# How to Teach a Teacher: Challenges and Opportunities in Physics Teacher Education in Germany and the USA

Ben Van Dusen, Christoph Vogelsang, Joseph Taylor, and Eva Cauet

**Abstract** Preparing future physics teachers for the demanding nature of their profession is an important and complex endeavor. Teacher education systems must provide a structure for the coherent professional development of prospective teachers. World- wide, physics teacher education is organized in different ways, but have to face similar challenges, like the relation between academic studies and practical preparation. To meet these challenges, it is worth taking look at different teacher education systems. In this chapter, we compare physics teacher education in two countries, representing two different educational traditions: Germany and the USA. Comparing different aspects of physics teacher education (standards, organization and institutionalization, content of teacher education, quality assurance), we describe both systems in their current state and why they are organized in the way they are. In doing so, we identify surprising commonalities but also different opportunities for both systems to learn from each other.

## Introduction

The goal of this paper is to inform physics teacher education by comparing two major traditions of physics teacher preparation. To compare different teacher education systems it is necessary to understand how and why they are designed as they are. Although every society has built unique educational systems grounded in a specific cultural context, scholars have identified two major educational traditions, which can be used to distinguish between systems on a general level: an Anglo-American tradition and a Continental-European tradition. They represent specific sets of paradigms and philosophies of teaching and learning (Sjöström et al., 2017), which also have an influence on the expected role and professional status of teachers and what and how teachers should learn during their preparation (Kansanen, 2009).

The Anglo-American educational tradition is assumed to be a major influence in the United States, United Kingdom, Australia, and other mostly english-speaking countries. The tradition has been described as a curriculum orientation with a primary goal of preparing students for the needs of society (Westbury, 2000). Curriculum in this sense is meant as a sequence of content and skills that are to be met in each year of instruction and are often referred to as "standards" in the U.S.. This corresponds to promoting scientific literacy, which focuses on the learning of science concepts for later application and its usefulness in life and society (Roberts,

2011). Fensham (2009) describes the role of teachers in this tradition as "agents of the system" (p. 1082), who are responsible for meeting a set of standards given by an external authority. This perspective also has an influence on how teacher education is designed. In a simplified way Tahirsylaj, Brezicha & Ikoma (2015) state, teacher education in this tradition has a strong emphasis on practical training.

The Continental-European educational tradition is assumed to be a major influence in the countries of Central and Northern Europe, especially in the German speaking nations and Scandinavia. Central to this tradition is the concept of Bildung, a German word, which can't be translated in one single term in English (Sjöström et al., 2017). One often cited description is pointed out by Westbury (2000): "Bildung is a noun meaning something like, being educated, educatedness'. […] Bildung is thus best translated as 'formation', implying both the forming of the personality into a unity as well as the product of this formation and the particular 'formedness' that is represented by the person" (p. 24). Bildung frames the emancipation of an individual as the overall purpose of education. Following an influential concept, three abilities should be fostered by Bildung: self-determination (being able to determine one's own personal life and interpretations of meaning in interpersonal, professional, ethical areas), co-determination (being able to take part in the development of society) and solidarity (with other members of society, especially when whose opportunities for self- and co-determination are limited) (Fischler, 2011). In this tradition, teachers are given a great amount of autonomy and are expected to transform knowledge/content in a way that it contributes to these goals (Fensham, 2009). Therefore, it's the task of teachers to build a curriculum in the above mentioned sense, roughly guided by more brief standards prescribed by an external authority. Teacher education in this tradition has a strong emphasis on theoretical studies of education, structured in specific sub-disciplines. Particularly, concepts to deal with the task of transformation of subject matter content for learning is reflected in the sub-discipline of Fachdidaktik, in the case of physics Physikdidaktik (Fischler, 2011).

We must be aware that these traditions are products of complex historical/philosophical developments, which contains a huge amount of simplification for sharp contrasting. They can't fully represent the whole complexity and richness of one teacher education system. To take a deeper look into the case of educating physics teachers in particular, we provide a comparative analysis of the teacher education systems for physics teachers in the United States and Germany. Both countries exemplify key aspects of these traditions. The aim of our analysis is to identify the strengths and potentials of both systems – leaving the reader with a broader view on the different ways that successful physics teacher education can be established. Furthermore, we are interested in examining to what extent these reconstructions still apply to the practice in the systems in their current state.

# Comparative framework

Based on the work of Blömeke (2008), Darling-Hammond (2017) and Tahirsylaj, Brezicha & Ikoma (2015) we developed a framework for comparing the educational systems. We start with comparing the *standards for teacher education* in Germany and the U.S. Then we describe how the identified differences manifest in the *organization of teacher education* with respect to entry to the profession and ongoing professional development. The heart of this chapter builds the comparison of the *contents of physics teacher education* which reflects country-specific emphasises regarding different knowledge areas. We also discuss the role of theoretical education and practical training in the different systems and how those elements are linked to each other. Finally, we compare the *quality assurance and control* of physics teacher education in Germany and the U.S.

## Standards for physics teacher education

What's the goal of physics teacher education? Every country has to specify the tasks teachers are expected to fulfil in their job and therefore the aims of their teacher education system.

### Standards for physics teacher education in Germany

Germany's 16 federal states have a high degree of autonomy in educational politics. Each federal state independently defines the objectives of schooling and consequently defines the goals of teacher education. Despite the differences, there are many similarities between the states and much is being done to harmonise systems while maintaining regional strengths.

The school system across the states can be characterised as follows. After completing a four to six year elementary school beginning in the age of six, students attend a secondary school out of three different tracks: One eight/nine year long academic track, the Gymnasium, leading to the highest possible school degree (*Abitur*), which allows students to attend university afterwards; one five year long track, the Realschule, for students seeking extended education, but do not wish to undertake an academic education; and one four year long track, the Hauptschule, focusing on preparing students for vocational training afterwards (i.e. learning a craft). In recent years more states have begun to integrate Hauptschule and Realschule into one track and also implemented comprehensive schools, leading effectively to two-track school systems. Future physics teachers are educated to be teachers for one to two of these tracks. Physics is taught as an mandatory, independent subject (as well as chemistry and biology) at all secondary schools (cf. DPG, 2016). Science as a comprehensive subject is only taught at some comprehensive schools.

In order to enable a certain degree of comparability, teacher education programs across the states are based on common standards formulated by the Standing Conference of the Ministers of Education and Cultural Affairs of the federal states (KMK). These standards exist for preparation in general educational sciences and also for each subject. The standards for educational sciences (KMK, 2019a) specify the professional competence teachers of all subjects and school types should achieve. They differentiate between four dimensions of competence: instruction, *Erziehen* (social and moral development & civil education), assessment and innovative development of schools. These dimensions formulate tasks, future teachers are expected to fulfil and cover a wide range of different topics, e.g. assessment approaches, teaching methods, socio-cultural influences on learning and so on. Another set of standards (KMK, 2019b) specifies the professional competence teachers should achieve for a specific subject. They also cover several aspects regarding physics education, but are formulated relatively briefly. For example, future physics teachers should have comprehensive content knowledge in physics, which enables them to design physics related learning environments in a professional manner. This implicitly reflects the expectation that teachers have the task of selecting content for teaching. Also the standards formulate a list of content of physics and physics education to be included in teacher education programs.

Based on these brief standards, every institution for teacher education is responsible for designing its own curriculum and specify goals in detail, leading to various programs differing between states and even between teacher education institutions within states. To ensure compliance with the standards, its programs have to be accredited by institutes, which are also accredited by a statewide accreditation council (Neumann, Härtig, Harms & Parchmann, 2017). In this process, teacher education programs are also examined, whether they are in line with the Bologna agreement of the European Union for harmonising the systems of higher education (Bauer et al., 2012).

### Standards for physics teacher education in the U.S.

Like in Germany, each of the 50 states and the District of Columbia have their own independent state and local school systems that create what has been described as a "sprawling landscape" (Cochran-Smith et al., 2016; Floden et al., 2020). U.S. Secondary schools are based on a liberal education model that eschew creating career tracks in favor of requiring students to take a breadth of course work. High school (grades 9-12) student graduation requirements are set by states but typically require 2-3 years of science classes that include at least 1 biological and 1 physical science course. A majority of high school students choose to take chemistry to fulfil their physical science courses. A 2013 survey found only 39% of students took any high school physics (White & Tesfaye, 2013). Of those 39% of students, 65% took a non-college credit earning physics course (e.g., physics 1st, conceptual or regular) and 35% took either honors or advanced placement physics.

Each state has its own teacher-certification department that sets the state's standards for K-12 education, teacher licensure, and grant accreditation to preservice teacher programs. Most teacher credential programs also receive external accreditation through a national organization called the Council for the Accreditation of Educator Preparation (CAEP, 2020). While there is a shared set of K-12 science education standards that states can choose to adopt (the Next Generation Science Standards; National Science Teaching Association, 2020) the effort to create a shared set of teacher licensure standards is not as well developed. The majority of states have content knowledge requirements that are measured through *Praxis* exams (ETS, 2020), although the specific scores and tests required vary by state. The education Teacher Performance Assessment (EdTPA; Sato, 2014) has also emerged as a more holistic assessment and support system but it is still only used in a minority (n=18) of states. Each preservice teacher program determines its own set of objectives for its graduates based on the state teacher performance expectations. While the state teacher performance standards vary by state, they have many attributes in common. California (the most populous U.S. state), for example, has six standards domains each with a set of more specific sub-standards: 1) engaging and supporting all students in learning, 2) Creating and maintaining effective environments for student learning, 3) understanding and organizing subject matter for student learning, 4) planning instruction and designing learning experiences for all students, 5) assessing student learning, and 6) developing as a professional educator (California Commission on Teacher Credentialing, 2016). Even within states, however, programs of teacher preparation vary in the size, duration, curriculum, and the nature of field experiences (NRC, 2010).

## Organisation and Institutionalisation of Teacher Education

Each teacher education system has specific pathways that prospective teachers typically have to follow in order to be able to work in this profession. In this section we describe these pathways for Germany and the U.S.

### How to become a physics teacher in Germany?

The typical pathway leading to the teaching profession in Germany has a relatively stable structure, which is similar for all federal states (Cortina & Thames, 2013). Teacher education is organised in two consecutive phases. The first phase of initial preparation involves studies at a university followed by a structured induction to the field at a particular school in the second phase. Further professional development is considered as the third phase of teacher education although this phase is not structured to the same extent as the first two phases.

### Initial Preparation

In the first step one has to enroll in a teacher education study program at a university. This requires a university entrance qualification, usually the *Abitur*. These programs are aligned to the school tracks (e.g., you can study for teaching at Gymnasiums). Preservice teachers also have to study two subjects taught at schools. Each university is responsible for designing its curricula autonomously.

Since the Bologna agreement of the EU, preservice teacher programs in the most federal states are organised in the Bachelor-Master-system. Students first have to obtain a Bachelor of Education (e.g., BA of Arts or Science) before earning a Master of Education. Most students acquire both degrees at the same university, since switching between universities is difficult and usually has disadvantages for the students because of the different curricula. Some federal states still organise their teacher education in the traditional study structure, in which students obtain a *first state examination* at the end of their studies without a degree in between. However, the length and scope of studies are comparable in both systems. A Master's degree or a first state examination is required to apply for the induction phase. Although the focus of the first phase is on the acquisition of more or less theoretical knowledge, typically several field experiences are integrated. The extent of these field experiences is usually defined by legal requirements of the federal states, so the scope and location in the course of a program vary between universities (Gröschner et al., 2015). Many programs include an initial orientation internship – typically a four week school placement – in the first two semesters, in which student teachers are supposed to reflect on their choice of teaching as a career, followed by one to two short school placements in the Bachelor's program. Eleven federal states have also implemented a one semester-internship at a school in their master degree programs (practical semester), which was accompanied by a shortening of the induction phase (Ulrich et al., 2020).

### Induction Phase

After finishing their studies, future physics teachers are entitled to apply for in-service training in the second phase, lasting from one and a half to two years. Formally, they undergo their training at a seminar for teacher preparation. These seminars are directly organised by the federal states, but each is responsible for designing their own preparation programs alongside the system of school tracks, which have to comply with the overall standards and state regulations.

Preparation takes place at two institutions. Most of the time, trainee teachers regularly teach a certain amount of classes at a school. They are usually mentored by experienced teachers from the same school, but in most states they also teach classes on their own with increasing amounts during their traineeship. In most federal states, every teacher at a school is expected to be able to serve as a mentor. Only very few states require them to undergo specific mentoring training and there is a discussion whether and how mentors need to be trained on a mandatory basis (e.g. Weylandt, 2012). Mentors are expected to provide feedback on the trainee teachers instruction and to support them in lesson planning and reflection. Parallel to teacher training at a school, the trainee teachers attend courses at the seminar. Courses are given by experienced teachers, who have passed an examination to serve as a teacher educator. Similar to their university studies the trainee teachers take courses on subject matter education and general educational studies, which focus more on practical training in comparison to the theoretical focus of the university studies. Coursework is usually one day a week.

During the traineeship the teacher educators regularly observe lessons of their trainees and grade their work. Also in many federal states mentor teachers (and sometimes the principals of the schools) have to provide short written reports on the professional development of the trainee teachers. At the end of the induction phase, the trainees have to undergo the second state examination. Elements of the examination differ in detail from state to state, but typically the trainees are required to present one examination lesson in each subject and take an oral exam on the course contents. Some states also demand a written thesis. Since much is at stake in this examination and the grades depend to a large extent on the subjective judgements of the examiners. there is always criticism of too intransparent grading criteria and unreliable assessment instruments (e.g. Strietholt & Terhart, 2009). After obtaining a second state exam, the trainees are fully licensed teachers and can apply to ministries or private schools for an appointment.

### Alternative pathways

Since there is a great shortage of qualified physics teachers in Germany too, many federal states provide alternative pathways for the entry in the profession. The requirements for this vary greatly and change from year to year, depending on the size of the shortage. Typically two pathways can be distinguished. In the first pathway, candidates with a masters degree of science corresponding to physics (e.g. physics, engineering, but also architecture etc.) can enter the induction phase. Some states require them to take a few courses in physics education and general educational studies at a university before or parallel. In the second pathway, teacher candidates with a masters degree are directly employed and work as teachers. Parallel they undergo on-the-job-training with the goal to get a second state examination. However, no federal state has made this second pathway entry accessible in recent years. Private schools are an exception, as they can decide on their staff independently. However, most students attend public secondary schools (ca. 91%, Klemm et al., 2018), private schools leading to secondary degrees are highly regulated and have to follow the same standards as public schools, so they often hire teachers with state licensure.

In the years 2002 to 2008 in Germany on average 45% of new physics teachers have entered the teaching profession following one of these alternative pathways, with an increasing proportion in the later years and differences between the school tracks (DPG, 2010). As researchers are only granted access to this data upon request, more recent results are not yet available. Evaluations indicate, teachers entering the induction phase without a Master of Education achieve similar levels of content knowledge for teaching, but less pedagogical knowledge, during traineeship (Oettinghaus, Lamprecht & Korneck, 2014). Several universities and organisations proposed programs for a better qualification for physics teachers entering the profession from the side (DPG, 2010).

Despite all these efforts, because of a shortage of physics teachers, in many federal states, teachers have to teach out of field to ensure the provision of physics lessons. Representative surveys reveal that roughly 6.5% of teachers are teaching physics out of their field (Stanat et al., 2019), with high variance between the federal states (between 1.7% and 17.9%). This undermines the strategy of quality assurance through high entrance qualification.

### How to become a physics teacher in the United States?

There is no coherent system for preparing physics teachers (Meltzer et al., 2012). The majority of physics teachers have neither a major or minor in physics. Furthermore, over 90% of physics teachers graduate from programs that do not offer any specialized instruction to prepare them for teaching physics. While 36% of physics departments report having a physics teacher education program, barely half of them report graduating any majors (Meltzer et al., 2012). If physics teachers are primarily not coming from physics or physics teacher education programs, where are they coming from?

Each year, there are around 3,100 new high-school physics teachers (White & Tesfaye, 2011). These 3,100 physics teachers come from two sources: 1) in-service teachers who are transitioning to teaching physics (n~1,700) and 2) first year teachers (n~1,400). The large number of in-service teachers transitioning to teaching physics is reflective of the severe shortage of physics teachers nationally. Many of these teachers transition from other science disciplines, while others are transitioning from unrelated disciplines. Both groups of teachers, however, are required to have earned state teaching licensure.

There are a range of paths to earning state teaching licensure and they vary by state. The paths are often described as either "traditional" or "alternative," but there is no commonly held agreement about how these are exactly defined (NRC, 2019).

### Traditional preparation

The traditional pathway for licensure is typified by the requirement of completing a teacher preparation program run by a university. These teacher preparation programs are usually 1-2 year long post-baccalaureate programs and offer students the opportunity to earn a masters degree in education while earning their licensure. Some states offer physics specific teaching certifications while others offer general science certifications that allow teachers to teach any science disciplines. Both types of certifications are likely to require some physics coursework to have been completed, but the requirements range from completing the introductory non-major sequence to completing a number of upper-division physics courses. While it is common for the teacher preparation programs to offer science teaching methods courses, it is very uncommon for them to offer any courses specific to physics teaching preparation. This is likely due to the small number of pre-service physics teachers in any given degree program making it impractical to offer coursework for them.

A central feature to most traditional licensure programs is a student teaching experience. Student teaching pairs pre-service teachers with one or more in-service teachers and provides an immersive teaching experience that ranges from weeks to months. In lieu of offering traditional courses during student teaching terms, preservice teacher programs typically will observe their pre-service teachers in the classroom and provide them feedback and support. These student teaching experiences are often the basis for capstone projects. Capstone projects are usually completed at the end of a licensure program and are meant to develop and demonstrate the breadth and depth of student knowledge in the field. The projects often include in depth reflection on and assessment of their capstone teaching. There have been attempts to create multi-state shared capstone expectations, such as edTPA (Sato, 2014).

The final component of most traditional teacher preparation programs is the completion of content specific *Praxis* exams. The specific requirements and exams vary by state. States that offer physics specific endorsements will typically require physics specific examinations to be completed while states that offer general science endorsements will typically require multidisciplinary examinations. Some programs, however, have been accredited in ways that allow them to offer examination waivers if students complete a specific set of courses.

Each state offers their own teaching license but once a person has received licensure from a state, other states offer forms of reciprocity that make it easier for them to earn licensure in them. State reciprocity programs often require passing of additional content assessments (Teacher certification degrees, 2020). The only way to receive licensure in all 50 states is to earn a national board certification (Goldhaber & Anthony, 2007). National board certification is only available to experienced teachers and employs a rigorous process of evaluating teacher quality.

### Alternative preparation

What qualifies as "alternative" preparation varies by state and while there are exceptions a common trait is that the programs are not run by 2- or 4-year colleges. The acute lack of teachers in some disciplines (e.g., physics) and regions have led many states to offer emergency credentials in high-need areas (Meltzer et al., 2012). Emergency credential standards vary but they typically drop any post-baccalaureate program requirements and focus on passing content assessments. Emergency credential programs often lack any formal training to develop pedagogical knowledge or pedagogical content knowledge.

Some teachers skip the licensure and credential processes all together by working in private or charter schools. Private and charter schools are not bound by many state standards and often employ teachers without licensure. While most students attend public schools, private and charter schools collectively enroll around 16% of the K-12 student population (Citylab, 2014; In perspective, 2018).

## Ongoing professional development

Teachers, teacher educators, policy makers as well as researchers - they all agree that ongoing professional development (PD) is an essential and necessary part of being a teacher. To be able to orientate themselves within the different PD programs future physics teachers need to know the main characteristics of high quality PD. Darling-Hammond, Hyler & Gardner (2017) identified seven criteria for effective professional development: "[Effective PD]
1. Is content focused
2. Incorporates active learning utilizing adult learning theory
3. Supports collaboration, typically in job-embedded contexts
4. Uses models and modeling of effective practice
5. Provides coaching and expert support
6. Offers opportunities for feedback and reflection
7. Is of sustained duration"

### Ongoing professional development in Germany

In Germany, there are two kinds of professional development programs: in-service training programs aiming to preserve and improve teachers' professional competences during their career and training programs that are required in order to apply for specific positions (e.g. headmaster positions, teacher educators in the induction phase) (cf. Eurydice, 2003). All federal states in Germany oblige their teachers to engage in professional development. While some expect their teachers to fulfil their obligations within their course free time others count at least a part of the invested time on teachers' teaching load.

Terhart (2000) differentiates between supply-led vs demand-led PD and school-intern vs. school-extern PD. Supply-led PD programs are normally offered by school-extern institutions, mostly by the "Landesinstitute" (institutions responsible for the quality assurance in school run by the federal states) but also by universities or organisations like the German Physical Society. Teachers can individually decide to participate in those programs if they are interested in the offered topics. School extern PD is the most common type of such programs in Germany. Their purpose is to engage teachers in content-specific learning processes absence of daily business. However, school-extern PD programs are often responsible for implementing administrative reforms, which are increasingly based on plausibilities rather than on scientific evidence (Pasternack et al., 2017). More teachers prefer demand-led programs, often realised within school intern PD programs, in which the staff of a school participates in so called pedagogical days, conclaves or supervisions, independent of the question if e.g. the collegium or extern referents organises and implements the events (Wenzel & Wesemann, 1990). School intern PD is mandatory in all federal states, but the specific obligations differ from state to state. It usually concentrates on the needs of the specific school, e.g. organisational development or teachers' personal professionalisation (e.g. method training, teacher health) (DVLfB, 2018).

Empirical data on German PD programs is scarce - and for physics teachers in particular. In a nation-wide survey of mathematics teachers and teachers of all science subjects in Germany (biology, chemistry, physics) (Richter, Kuhl, Haag, and Pant, 2013), 85% of physics teachers reported to have participated in at least one PD within the last 2 years, while 15% of the teachers did not attend any PD program. With regard to contents, PD programs most frequently attended by physics teachers (25%) were focused on how to impart physics topics in classroom setting (pedagogical content knowledge) followed by programs on subjects unspecific forms of teaching and methods (pedagogical knowledge) (20%). Teachers at the Gymnasium participate significantly more often in PCK or CK related PD than teachers teaching at the other school tracks while the picture looks the opposite for PK related PD programs. Teachers who did not participate in PD during the last two years reported organisational barriers (time conflicts 72%, difficulties to find substitutes for their classes 53%) but 40-50% also reported little practical benefit, or disappointing experiences from former PD participation (c.f. Krille, 2020).

### Ongoing professional development in the U.S.

In the U.S., K-12 teachers have a vast array of professional development opportunities. While the large library of options is inherently a positive characteristic, what is often associated with such volume is a lack of systematicity and coherence (NRC, 2020). However, it is useful to map the landscape of PD in the U.S. by describing themes in delivery formats, teachers' time participating, content foci, and alignment of activities with principles of effective PD. With regard to describing themes in professional development for science teachers, and physics teachers in particular, the most current and comprehensive source of empirical data is the nationally representative 2018 National Survey of Science and Mathematics Education (2018 NSSME+; Banilower et al., 2018). In the following section, we provide selected findings from this survey.

Among the many available delivery options, science teachers in the U.S. most often participate in PD via a workshop format. Subject-specific PD is often a part of what science teachers participate in with approximately 80 percent participating in science-specific PD in the last 3 years. However, the quantity was modest with only about one third of high school science teachers participating in more than 35 hours of subject-specific professional development in that three-year period.

Science teacher respondents to the NSSME+ indicated that the alignment of their PD experiences with the elements of effective PD was moderate (average score of about 50 on a 100-point alignment scale) where the elements of effective PD included having teachers work with colleagues who face similar challenges, engaging teachers in investigations, and examining student work/classroom artifacts. The 2018 NSSME+ results also indicate that 63% of physics teachers participated in science-specific PD in the last year and 85% had participated in such in the last three years.

In terms of topical coverage, the 2018 NSSME+ collected teachers' ratings of the extent to which their PD offerings emphasized selected topics. From those data, the combined percentage of teachers rating each topic as a 4 or 5 on a 5-point emphasis scale can be used to rank the topics on perceived emphasis. In terms of highlighting the most emphasized topics, the combined percentage of teachers giving a topic an emphasis score of 4 or greater was 54% for deepening understanding of how science is done, 43% for monitoring student understanding, 42% for developing science content knowledge, 38% for differentiating instruction, and 38% for integrating STEM content.

The 2018 NSSME+ also provided data on how teachers were engaged in PD and what opportunities teachers they had during PD. From those data, we provide the combined percentage of teachers rating each PD opportunity as a 4 or 5 on a 5-point "extent of opportunity" scale. In terms of highlighting the most often provided learning or collaboration opportunities, the combined percentage of teachers giving an opportunity an extent of opportunity score of 4 or greater was 51% for working with other teachers of the same subject or grade level, 49% for working with other teachers from the same school, and 47% for engaging in scientific investigations or engineering design challenges.

## Content of Teacher Education

Main goal of teacher education systems is to foster future teachers' development of professional knowledge and skills. There are many models of the professional knowledge base for teaching physics. We use the Refined Consensus Model of Pedagogical Content Knowledge (PCK) (Carlson & Daehler, 2019) to give a comparative overview of the contents of teacher education in the U.S. and Germany. The model describes the interplay of different kinds of teacher knowledge. First, it distinguishes several professional knowledge bases: content knowledge, pedagogical knowledge, knowledge of students, curricular knowledge and assessment knowledge. These knowledge bases contribute to the collective PCK, in short the knowledge of how to teach a physics topic or concept in an appropriate way. The individual knowledge of a particular teacher is referred to as personal PCK, whereas enacted PCK describes „the specific knowledge and skills utilised by an individual teacher in a particular setting" (Carlson & Daehler, 2019, 84)

### Content of teacher education in Germany

German teacher education programs often formulate their goals alongside models of professional competence, which integrate models of professional knowledge (Baumert & Kunter, 2013). From an overarching perspective, most programs structure their curriculum into three knowledge areas: content knowledge of physics, knowledge in physics education (*Fachdidaktik*, Fischler, 2011) and knowledge of general educational concepts. These areas can be found in all phases of german teacher education, but their scope and proportion change during the path of preparation.

In their university studies future physics teachers mainly have to take courses focusing on physics content knowledge. In terms of the European Credit Transfer and Accumulation System (ECTS) a study program for teacher students has to cover 300 credit points. One point represents a study work load of 25 to 30 hours. A typical study program for physics teachers includes 90 ECTS-points for subject matter studies and 30 ECTS-points for studies in physics education (40 credit points accounts for general educational studies, 120 points for the second subject) (DPG, 2014). Proportions differ between programs focusing on different school tracks. Content courses typically reflect broad studies in different areas of physics as defined in the standards (KMK, 2019). Regarding experimental physics, this includes lessons in mechanics, thermodynamics, electricity, optics, atom and quantum physics (Neumann, Härtig, Harms & Parchmann, 2017). Level and depth of the studies also vary between the study programs. Students studying for the tracks of Haupt- and Realschule also take basic classes on solid state, nuclear and particle physics, students for the Gymnasium are expected to gain deeper knowledge in these areas. Regarding theoretical physics, students for Gymnasium have to take lessons on theoretical mechanics, thermodynamics, electrodynamics and quantum mechanics. Students for Haupt- and Realschule in comparison are expected only to obtain a basic overview of the structure and main concepts of theoretical physics. All students have to take basic lab work courses and courses on school-oriented experimentation, students for Gymnasium also take advanced lab work courses. Students for all tracks have courses on applied physics, leading to an overview of relevant topics for schooling (e.g. climate and weather, physics and sport). Also, students are expected to learn aspects of the nature of physics. At most universities, teacher students usually take courses together with students studying physics full-time, but most lecturers don't see themselves as teacher educators and prepare their courses mostly not for teacher students. Another important factor is, that problems with coping with content studies are one of the main reasons for dropouts (cf. Heublein & Schmelzer, 2018). Regarding physics education, courses cover physics education theories and conceptions, students' motivation and interest, learning processes, learning difficulties and students conceptions' of physical concepts, use of experiments, lesson planning and reflection on physics instruction, use of digital media in instruction, heterogeneity of students and topics of recent physics education research. Amount of courses also vary depending on the focused school track. These courses contribute to the knowledge of students and provide collective PCK. Curricular knowledge and assessment knowledge are blind spots in german teacher study programs, as they differ greatly in this respect. Also, how courses are structured is highly variable between universities. In summary, the overall approach of the first phase is a great emphasis on learning rather theoretical knowledge and looking at physics instruction from the perspective of theory. In recent years a lot of research was carried out, to evaluate whether students acquire this knowledge like expected. Most studies find evidence for positive development of content knowledge and PCK in general. In detail deficits were identified regarding specific aspects, like great differences in content knowledge and PCK between students studying for different tracks (e.g. Riese & Reinhold, 2012).

The following induction phase focuses more on practical teacher training. The overall approach is similar to the concept of the reflective practitioner by Donald Schön (1984). The trainee teachers are expected to apply their theoretical knowledge in actual classroom instruction and also use it to reflect on their teaching (and the teaching of others). In terms of the model, in the induction phase future teachers develop mostly personal PCK and reflect on their enacted PCK. The content covered in the complementary courses during traineeship contributes to this, by focusing on practical issues on how to deal with concrete, specific tasks the trainees have to cope with at their schools. Many programs also include curricular knowledge and knowledge of assessment, but usually with a great focus on practical demands. Also course content on regulatory and school-law issues is part of the curriculum in most programs. Similar to the first phase, how content is taught, varies greatly between different teacher preparation seminars. Compared to the first phase, there are fewer evaluations of the effectiveness of the induction phase (e.g. Plöger, Scholl. Schüle & Seifert, 2019), especially with a focus on physics.

In german teacher education, too, the theory-practice-gap is a major challenge. In the domain of physics, in Germany low to no correlations between teachers' professional knowledge, the quality of their instruction and/or achievement of their students could be found (e.g. Cauet et al., 2015). In terms of the Consensus Model only few relations between the knowledge bases or personal PCK and enacted PCK were observed. The implementation of one-semester-internships into master study programs is a reaction to this and the attempt to make more connections between theory and practice possible while students are still at university (Ulrich et al., 2020).

### Content of teacher education in the U.S.

The content of teacher education ranges from comprehensive traditional education programs to alternative education programs that offer no training. For the purposes of this section, we will focus on the traditional education programs. Traditional teacher education programs are post-baccalaureate programs that assume that their preservice teachers gained their content knowledge as part of their undergraduate coursework (NRC, 2019). The prior physics coursework ranges from completing a pair of introductory physics courses to a traditional bachelor's degree in physics. While 36% of physics departments have a physics teacher education program, barely half of them are actively graduating students (Meltzer et al., 2012). Nationally, there are annually only around 270 students who graduate from a physics teacher education undergraduate program in either a physics department or a school of education (Meltzer et al., 2012). This means that only 8.7% of the 3,100 first-year physics teachers earned bachelor degrees that explicitly developed physics PCK.

The post-baccalaureate teaching licensure programs assume that students have sufficient content knowledge and therefore focus on developing pedagogical knowledge and PCK. The coursework is a mix of graduate-level general education and science teaching courses. Typical course credit requirements total around 35 semester credits with about ⅓ of those being science teaching specific credits. Common general education course topics include educational theory (e.g. constructivism, multiple intelligences, and zones of proximal development), U.S. educational history (e.g., normal schools, school integration, and the accountability movement), education law (Brown v. Board, compulsory education, and teacher/student rights), educational technology (e.g., assistive technology, remote learning, and asynchronous learning), educating diverse student populations and social justice (e.g., critical self-reflection and equitable pedagogical practices).

Science teaching courses are designed to develop general science PCK and are taken by a blend of pre-service teachers across the science disciplines. Despite typically being required to be taken for multiple terms in a program, it is very rare for a course to teach physics specific PCK. The lack of physics PCK is likely due to the very small number of pre-service physics teachers in a program in any given year. Common science teaching courses topics include the nature of science, research on effective science teaching, creating equitable science learning outcomes, fostering productive science talk, creating lesson plans that meet state science standards, and making science relevant to students. These courses often use a book as a central organizing artifact (e.g., Ambitious Science Teaching; Windschitl et al., 2018).

The physics specific PCK is primarily developed through student teaching. Student teaching pairs pre-service teachers with in-service teachers where they spend several months apprenticing in secondary school science courses. While student teaching, the pre-service teacher leads several units of instruction with oversight and support of the in-service teacher and a university supervisor. It is common for pre-service teachers to find a lack of coherence between their highly theoretical coursework and their real-world student teaching experiences (Zeichner, 2010).

## Quality Assurance and Control

Governments try to ensure the quality of their future teachers and their work using various strategies. Regarding the selection and recruitment of future teachers governments can, for example, manage the total number of places available for teacher education students, try to influence the attractiveness and status of teaching as a profession and a career and specify the requirements and qualifications needed, to enter the profession (Ingvarson & Rowley, 2017).

Some of these quality assurance arrangements relate to working conditions of teachers, while others refer more to regulations of teacher education programs and standards. Many countries have also implemented quality control measures in order to improve the work of in-service teachers.

**Quality Assurance and Control in German Teacher Education**

The attractiveness of teaching as a career depends on various aspects like status, working conditions and salary scales. In Germany, teachers generally can have one of two different employment statuses. Most teachers are civil servants with lifetime tenure. They are working under a distinct regulatory framework (Eurydice, 2003). and must follow a specific code of conduct/ethics. This status includes some privileges like special health care support and state backed pension plans. On the other hand, civil servants aren't allowed to go on strike and cannot choose totally free at which schools they will work. The minority of teachers have the status of employees, meaning they are employed on a contractual basis following general employment law. Most of them work under permanent contracts, temporary contracts are exceptions offered mainly to substitute teachers, who are on sick or parental leave. Regardless of status, most teachers are employed directly by the federal state. For private school, the respective school board is the employer. Overall, the job security of employed teachers is extremely high resulting in a strong professional identity of teachers as officers of the state (Eurydice, 2003). Salaries are based on collective bargaining and salaries of teachers as employees are often a little lower than that of teachers as civil servants, although work requirements are the same. Against the background of a general shortage of teachers, 15 out of 16 federal states are employing new teachers as civil servants. Salary scales of secondary teachers on average are comparable to other professions requiring a master degree in Germany (Ingvarson & Rowley, 2017). However, once employed, there are few opportunities for promotion or further career paths: teachers can become principals (requiring further training), teacher educators or take further duties in school, for example maintaining the computer lab in order to improve their pay grade. Another factor influencing the attractiveness of teaching as a career are conditions of teacher preparation and in the workplace. At all public universities in Germany students do not have to pay tuition fees – they simply have to cover their living expenses by themselves. In the induction phase they even receive a reduced salary. Regardless of their status, teachers have to teach a mandatory amount of 23 to 27 lessons (45 minutes) a week, depending on the school track and the federal state (KMK, 2017).

Since teaching as a profession is quite attractive, there are concerns, that entrants in teacher education programs are less qualified and – for example - have lower

GPAs (*Abiturnote*) than entrants in other programs, but studies show no evidence for this assumption in general (Henoch, Klusmann, Lüdtke & Trautwein, 2015). However, regarding future physics teachers, students for the Gymnasium-track have higher GPAs and begin their studies with a higher level of prior mathematics and physics knowledge than students for the other tracks (c.f. Riese & Reinhold, 2012).

The number of places available for teacher education students is mostly regulated by the governments of the federal states. Because of the low number of enrolments, study programs for physics teachers do not have to use any procedure of selection. Drop-out rates in study programs in physics at German universities are relatively high between 30% and 40% (Heublein et al., 2018).

As described in the previous section, following the ideal pathway, one must complete a long and highly structured qualification phase to be entitled to apply for an appointment as a teacher. However, once in-service little further professional development or certification is required/mandatory. For a few activities physics teachers must obtain additional certificates, for example to be allowed to show experiments regarding radioactivity in the classroom. Although teachers are obliged to engage in professional development, only 3 out of 16 federal states formulate verifiable criteria for professional development by quantifying the amount of trainings they expect teachers to absolve (12x5 hours within 4 years i.e. 15 hours/year in Bavaria, 30 hours/year in Hamburg and Bremen). Only 9 federal states insist on explicit documentation (e.g. in a portfolio) of how teachers fulfil their PD obligations and even less expect headmasters to use it as a base for individual career development during e.g. annual performance reviews (DVLfB, 2018, p.20). For Germany as a whole, there are neither nationwide standards for assuring the quality of professional development programs, nor do nationwide monitoring, evaluations or reporting exist in order to gather data for quality controls (DVLfB, 2018, p.124). Even worse, only some federal states require governmental approvement of PD offers - and in most cases those are based on self-declarations of the PD suppliers and only require the adhesion of formal minimum standards (e.g. information on content and didactic and methodical design as well as on the acquirable competencies has to be provided, and programs have to fit to the school law and it aims) (Pasternack et al., 2017, p.68, 287).

The responsibility for school supervision lies with the individual states. Following an evidence-based approach of school governance, in recent years all federal states have implemented some kind of school inspection as part of quality control measurements (Altrichter & Kemethofer, 2016). The typical process is similar in all states. One school is inspected once every several years (roughly five years). External inspectors visit the school for several days, observe classes, interview several stakeholders (principal, staff, parents, students), collect information on some school aspects provided by the school (e.g. management plans) and write an inspection report based on statewide standards. This report is given back to the school and it is expected that the respective school will take it as a starting point for quality developments. On a national level, the Institute for Educational Quality Improvement (IQB) is responsible for providing the federal states with information for school development and for monitoring the extent to which Germany's students are achieving educational standards. Therefore, the IQB is carrying out nationwide assessments based on representative samples of schools and students every one or two years. The IQB reports the results of these assessments to the governments of the states, to the participating schools and in some cases to teachers and students. However, assessments regarding physics are seldom carried out (last in 2012 and 2018, e.g. Stanat et al., 2019) and these assessments have no direct consequences for individual teachers, neither positive nor negative.

### Quality Assurance and Control in the U.S.

In the U.S., states typically require teacher candidates to pass an assessment of relevant content knowledge, pedagogical knowledge, and/or basic skills (.e.g, reading, writing, mathematics). The *Praxis* tests (ETS, 2020) are required for licensure in more than 40 U.S. states although the requirements for test content varies. For example, in some states, prospective physics teachers are required to pass a *Praxis General Science Content Knowledge* test, while in other states, passing the *Praxis Physics Content Knowledge* test is required. This variation is due in part to the fact that licensing specificity differs across states, some states offering only a secondary school science certification, while others certify prospective teachers specifically in physics.

Once a prospective teacher is hired for their first appointment in a public school, they usually become employees of a geographically defined school district, but may draw retirement or other benefits through state-based programs. Often, especially in rural areas, school district boundaries are synonymous with county boundaries. In many states, teacher salaries are based on a collective bargaining agreement between the local teacher union and the school district board of directors. Collective bargaining agreements usually include salary schedules where one's salary is jointly determined by experience, graduate degrees obtained, and/or graduate credits earned.

On the professional development front, newly hired physics teachers may have access to a teacher induction program which often includes being assigned a local mentor. According to (NRC, 2020), approximately 70 percent of U.S. schools offer formal teacher induction programs, with a third of those schools offering programs lasting one year or less, and very few offering induction programs three years or longer. When induction programs exist, they are most often developed locally by the school or school district. This is somewhat consistent with previous work by Goldrick, Osta, Barlin, and Burn (2012) who found that 27 states require some form of induction support for new teachers, with 11 requiring two or more years of support. Further, 22 states were found to require completion of an induction program for an advanced teaching certification. In the U.S., some states offer two levels of certification, one is provisional/probationary that is concurrent with enrollment in an induction program and the second is received upon completion of an induction program and an initial demonstration of teaching effectiveness. The second certification can also coincide with professional tenure. In subsequent years, U.S. teachers do not have any federal requirements for ongoing professional development and license renewal but all individual states have such requirements and these vary significantly across states in the U.S..

Evaluation of teacher effectiveness for certification and other purposes is determined primarily by individual districts and states with flexible national input via the Every Student Succeeds Act (ESSA, 2015)) which provides high-level guidance that teacher effectiveness ratings be at least in part derived using evidence of student growth. The degree to which student growth on state tests is a factor in teacher evaluation, and the choice of other evaluative factors is left entirely up to individual states and districts.

In 2019, only about one half of U.S. states required annual teacher evaluations (Ross&Walsh, 2019). When teachers are evaluated, a primary source of effectiveness evidence, beyond student outcomes, is teaching observations. Citing research on the unreliability of a single observation for capturing a teacher's overall effectiveness, most U.S. states require multiple observations of teachers within a given evaluation period, and those observations are made by any combination of colleagues, school administrators, third-party evaluators, and the teachers themselves. With regard to the larger school context, the Every Student Succeeds Act holds schools accountable for growth in student achievement but allows the parent school district to design and implement their own customized plans for correcting the course of low-performing schools.

## Discussion

In this section we provide brief overviews of the teacher education systems of Germany and the United States. We will compare these systems in order to identify each system's strengths and potentials, in order to give an insight into the different ways that physics teacher education can be designed,and how approaches borrowed from one county/tradition may help address the challenges of the other. Also, country comparisons can lead to a deeper knowledge about fundamental cultural concepts behind educational features (Blömeke & Paine, 2008, 2030). Therefore, we chose to describe and compare these two countries, as they were identified by several scholars, using information available at the time, as representatives of two leading educacional traditions (cite). In our comparison we attempt to identify observed differences across these traditions, noting that these models may no longer be adequate .

Looking onto the school system in general one major difference between Germany and the U.S. is that physics is a mandatory subject for all students in secondary schools in Germany, compared to a system with more options for course choices in U.S. K-12 Education. This can be traced back to the underlying concept of Bildung in this educational tradition, in which it is assumed, every student should have an essential level of physics knowledge to become a self-determined citizen. This difference is also reflected in the teacher education system. Future teachers in Germany are prepared as teachers for physics. Multi-disciplinary science teachers are an exception in Germany, whereas teachers who are prepared specifically to teach physics are less common in the U.S.. Both countries are similar in that the federal states have a great amount of autonomy regarding educational policies. Therefore, both countries have, in fact, a diversity of teacher education throughout the country. However, in Germany the states have agreed upon a set of standards for teacher education in general and particular for physics all teacher education programs have to be designed accordingly. In the U.S. a shared set of standards has not yet been implemented to the same extent. However, the German standards were also formulated only recently (considering the history of the German education system) as a reaction to the results of international students assessments like TIMSS and PISA. Thus, they are not rooted in the German educational tradition and more an example for the adjustment of german teacher education to ideas from other educational systems.

Also, there are differences between Germany and the U.S. regarding the preparation programs of future physics teachers. In the traditional pathway in Germany future physics teachers are enrolling in teacher education programs right from the beginning, developing PCK and PK in their undergraduate studies. This corresponds to the underlying educational tradition with its emphasis on theoretical studies in specialised sub-disciplines like Physikdidaktik (Fischler, 2011). Although U.S. universities also offer undergraduate physics teacher preparation programs most programs are post-baccalaureate and are not focused on physics. However, this makes it easier for students with science related degrees to switch to the teaching profession. Also, CK specific courses are less common in U.S. post-baccalaureate programs than in german teacher education programs. The teaching experiences in U.S. programs are similar to the practical semester implemented in most federal states in Germany and have elements similar to the german induction phase (like the capstone projects). On the other hand, a structured induction phase organised by the states is an obligatory component in the german preparation. In the U.S. system induction phases are part of teacher preparation, but are more locally based and not mandatory in all states. In the U.S., there are opportunities to advance on a salary scale through additional professional development and coursework, which is not common in Germany. The overall length of teacher education in both countries, however, is relatively similar. The content of teacher education is also reasonably similar. It is difficult to describe general differences in detail, due to the variance of programs between the states and universities in both countries.

Both countries suffer from a shortage of physics teachers and offer alternative preparation. In all federal states in Germany the minimum requirement is to complete the induction phase for an alternative entry into the profession.Most U.S. states offer emergency credentials that lack any formal teacher training. On the other hand, professional development could be regarded as the blind spot of the german teacher education system (DVLfB, 2018). Compared to the U.S. the options for PD are smaller and teachers in the US are taking longer courses. In both countries, teachers prefer to attend subject-specific PD and requirements for teachers to attend PD are varying between the states. From an overall perspective, the german strategy is to ensure quality of pre-service teacher education with great requirements for the entrance into the profession. The U.S. strategy is more focused on in-service professionalisation which is somewhat forgotten in Germany.

Although both countries stand as prototypes for two educational traditions based on different philosophies of education, in practice there are many commonalities. The requirements for adequate preparation of future physics teachers seem to have more influence on teacher education than the educational tradition of each country, at least in recent decades. Especially since the 2000s both countries are influenced by international large scale assessments for students achievement and developed similar approaches (like the development of new standards). Also in physics teacher preparation both countries have to face similar challenges. Both Germany and the U.S. have to deal with a shortage of physics teachers and the small number of students enrolling in teacher preparation programs. In both countries, a significant number of teachers teach physics out of field. Both countries have to cope with the gap between theoretical studies and practical demands of instruction.

Despite all these commonalities, neither country seems to have found an all-encompassing solution to these problems. Therefore, a strategy for further research could be comparative analyses, taking also high-achieving countries in assessments like PISA and TIMSS into account. But it would also be promising to examine differences between physics teacher education in Germany and in the U.S. more closely at the level of concrete program design than is possible in this short chapter. By looking at the details, further ideas and possibilities for improving one's own teacher preparation programs could be gained.

\*\*\*\*\*\*\*\*\*\*\*\*\*\*\*\*\*\*\*\*\*\*\*\*\*\*\*\*\*\*\*\*\*\*\*\*\*\*\*\*\*\*\*\*\*\*\*\*\*\*\*\*\*\*